\documentclass[
    ,final            
  ]{aipproc}

\layoutstyle{6x9}

\begin{document}
\newcommand{\be}{\begin{equation}}
\newcommand{\ee}{\end{equation}}
\newcommand{\bea}{\begin{eqnarray}}
\newcommand{\eea}{\end{eqnarray}}
\newcommand{\nn}{\nonumber}

\title{Polarization  in charmless $B \to VV $ decays}

\author{Fulvia De Fazio}{
  address={Istituto Nazionale di Fisica Nucleare, Sezione di Bari, Italy}
}

\begin{abstract}
 Recent data for $B$ decays to two light vector mesons show
 that the longitudinal
amplitude dominates in $B^0 \to \rho^+ \rho^-$,  $B^+ \to \rho^+
\rho^0$, $\rho^0 K^{*+}$ decays and not in   $B^0 \to \phi
K^{*0}$, $B^+ \to \phi K^{*+}$. We consider rescattering mediated
by charmed resonances, finding that in $B \to \phi K^*$  it can be
responsible of the suppression of the longitudinal amplitude.   A
similar result is found for $B \to \rho K^{*}$.
\end{abstract}

\maketitle


  Recently,
 decay widths and   polarization fractions of several B decays to
two light vector mesons were measured \cite{belle,babar}. The
 branching ratios are of ${\cal O}(10^{-5})$. The measured
polarization fractions, collected in  Table~\ref{tab:tab2}, show
that in penguin induced  $B \to \phi K^*$ transitions the
longitudinal amplitude does not dominate. However, using
factorization and  the heavy quark limit one can show that the
light VV final state should be mainly longitudinally polarized.
Actually, the decay $B^0 \to \phi K^{*0}$ is described by the
amplitude $ {\cal A}(B^0(p) \to \phi(q,\epsilon) K^{*0}(p^\prime,
\eta))= {\cal A}_0 \, \epsilon^*\cdot \eta^* + {\cal A}_2 \,
(\epsilon^*\cdot p) (\eta^* \cdot q)+ i {\cal A}_1 \,
\epsilon^{\alpha \beta \gamma \delta} \epsilon^*_\alpha
\eta^*_\beta p_\gamma p^\prime_\delta $, with
$\epsilon(q,\lambda)$, $\eta(p^\prime,\lambda)$ the $\phi$, $K^*$
polarization vectors and  $\lambda=0,\pm1$ the helicities. Since
the  $B$ meson is spinless, the final  mesons have the same
helicity.  In terms of ${\cal A}_{0,1,2}$ (describing  S, P, D
wave decays, respectively) the helicity amplitudes ${\cal A}_L$,
${\cal A}_\pm$ read: ${\cal A}_L=-  [ (p \cdot p^\prime-M^2_{K^*})
{\cal A}_0+M_B^2 |\vec p^\prime|^2 {\cal A}_2]/(M_\phi M_{K^*})$,
${\cal A}_\pm =- {\cal A}_0\mp M_B |\vec p^\prime| {\cal A}_1$.
The transverse amplitudes are defined as ${\cal
A}_{\parallel,\perp} =\displaystyle{({\cal A}_+ \pm {\cal A}_- )/
\sqrt 2}$, while
 the polarization fractions are
$f_i=\displaystyle{|{\cal A}_i|^2 / |{\cal A}|^2}$,
$i=L,\parallel,\perp$.

Considering  the effective weak Hamiltonian inducing the $\bar b
\to \bar s s \bar s$ transitions \cite{Ali:1998eb} the amplitude
${\cal A}(B^0 \to \phi K^{*0})$  admits a factorized form $ {\cal
A}_{fact}(B^0 \to \phi K^{*0})= -(G_F / \sqrt 2)V^*_{tb} V_{ts}
a_W \langle K^{*0}(p^\prime, \eta)|(\bar b s)_{V-A}|B^0(p)\rangle
\langle \phi(q,\epsilon)|(\bar s s)_V|0 \rangle $, with $a_W$  a
combination of Wilson coefficients. Using $\langle \phi(q,
\epsilon)|\bar s \gamma^\mu s|0 \rangle = f_\phi M_\phi
\epsilon^{*\mu}$ and \bea \langle K^*(p^\prime, \eta)|(\bar b
 s)_{V-A}|B(p) \rangle = - i \epsilon_{\mu \nu
\rho \sigma} \eta^{*\nu} p^\rho p^{\prime \sigma} {2 V(q^2) \over
M_B + M_{K^*}}  - \big [ (M_B + M_{K^*}) A_1(q^2) \eta^*_\mu
\nn\label{eq:BK} \\
 - {A_2(q^2)  \over M_B + M_{K^*}} (\eta^* \cdot p) \, (p +
p^\prime)_\mu - 2 M_{K^*} {(A_3(q^2)-A_0(q^2)) \over q^2}
(\eta^*\cdot p) q_\mu \big] \,\,,  \label{eq:formf} \eea
($q=p-p^\prime$), one can write the polarization fractions and
check that, for large $M_B$: ${\cal A}_L \propto M_B^3 [(
A_1(M_\phi^2)-A_2(M_\phi^2))+{M_{K^*}\over M_B} (
A_1(M_\phi^2)+A_2(M_\phi^2))]$, $ {\cal A}_\parallel \propto M_B
A_1(M_\phi^2)$ , ${\cal A}_\perp \propto M_B V(M_\phi^2)$. For
$M_B \to \infty$,  $q^2=0$ it was found that: $A_2/A_1=V/A_1=1$
\cite{Charles:1998dr}, giving $f_L \simeq 1 + {\cal O}(
M_B^{-2})$,
 $f_\parallel/f_\perp\simeq 1$. Using
generalized factorization,  considering  the $a_i$ as effective
parameters, one may reproduce the experimental branching ratio,
but not the polarization fractions, since the dependence on the
$a_i$ cancels in the ratios. Hence, to explain the small $f_L $
one has to look either at finite mass corrections and effects
beyond factorization, or at new physics \cite{Grossman:2003qi}.
Here we present the study
 in \cite{Colangelo:2004rd}. Since,
 using form factors computed for finite heavy quark mass,
 the  polarization fractions in
$B^0 \to \phi K^{*0}$
 do not fit data, in \cite{Colangelo:2004rd} we considered
 rescattering of intermediate charm states,
  already studied in \cite{Colangelo:1989gi}-\cite{chic0} and
  more recently in \cite{cheng}, showing that they
can invalidate the dominance of the longitudinal configuration in
 $B \to \phi K^*$  without affecting the observed $B
\to \rho \rho$ modes.

\begin{table}
\begin{tabular}{ccccc}
\hline \tablehead{1}{r}{b}{Mode}
  & \tablehead{1}{r}{b}{Pol. fraction}
  & \tablehead{1}{r}{b}{Belle~\cite{belle}}
  & \tablehead{1}{r}{b}{BaBar~\cite{babar}}
  & \tablehead{1}{r}{b}{Average}   \\
\hline
$B^+\to\phi K^{*+}$&$f_L$&$0.49\pm0.13\pm0.05$&$0.46\pm0.12\pm0.03$&$0.47\pm0.09$\\
                  &$f_\perp$&$0.12^{+0.11}_{-0.08}\pm0.03$& &\\
$B^0\to\phi K^{*0}$&$f_L$&$0.52\pm0.07\pm0.05$&$0.52\pm0.05\pm0.02$&$0.52\pm0.04$\\
                   &$f_\perp$&$0.30\pm0.07\pm0.03$&$0.22\pm0.05\pm0.02$&$0.24\pm0.04$\\
\hline
$B^+\to\rho^0 K^{*+}$&$f_L$&&$0.96^{+0.04}_{-0.15}\pm0.04$&\\
$B^+\to\rho^0 \rho^+$&$f_L$&$0.95\pm0.11\pm0.02$&$0.97^{+0.03}_{-0.07}\pm0.04$&$0.96\pm0.07$\\
$B^0\to\rho^+ \rho^-$&$f_L$&&$0.98^{+0.02}_{-0.08}\pm0.03$&\\
\hline
\end{tabular}
\caption{Polarization fractions in charmless $B\to VV$
transitions.} \label{tab:tab2}
\end{table}
The  decay $B^0 \to \phi K^{*0}$ can also be induced by
rescattering through
 $B \to D^{(*)}_s D^{(*)} \to \phi K^*$. Such effects could be sizeable
since they involve  Wilson coefficients of current-current
operators  (${\cal O}(1)$), while the  coefficients of penguin
operators in
 $\bar b \to \bar s s \bar s $ are ${\cal
O}(10^{-2})$. Besides, there is no CKM suppression  ($|V^*_{tb}
V_{ts}|\simeq|V^*_{cb} V_{cs}|$).
 The vertices
$D^{(*)}_s D^{(*)} K^*$, $D^{(*)}_s D^{(*)}_s \phi$ can be
estimated using an effective Lagrangian describing the
interactions of heavy hadrons with light vector mesons
\cite{Casalbuoni:1992gi}. We write $ \langle D_s^{(*)+} D^{(*)-} |
H_W | B^- \rangle = \displaystyle{(G_F/ \sqrt{2})}V_{cb}V_{cs}^*
a_1 \langle D^{(*)-} | (V-A)^\mu | B^0 \rangle \langle D_s^{(*)+}|
(V-A)_\mu | 0 \rangle $, with $a_1\simeq 1$, as there is empirical
evidence that factorization works in these modes. In the heavy
quark limit  the above matrix elements involve the Isgur-Wise
function $\xi$ and  a single quantity $f_{D_s}=f_{D^*_s}$; we use
$\xi(y)= \left(2/ (1+y) \right)^2$ and $f_{D_s}=240$ MeV.

Since the exchanged  mesons are off-shell, we write the couplings
$ g_i(t)=g_{i0}\,F(t)$, $g_{i0}$ being  on-shell couplings and
$F(t)=(\Lambda^2 -M^2_{D^*_s})/ (\Lambda^2-t)$ to satisfy QCD
counting rules. The relative sign of rescattering and factorized
amplitude is  unknown, as well as the role of diagrams involving
excitations, hence we fix $\Lambda=2.3$ GeV  analyzing the sum
${\cal A}={\cal A}_{fact} + r {\cal A}_{resc} $ in terms  of the
parameter $r$.
\begin{figure}\includegraphics[height=.25\textheight]{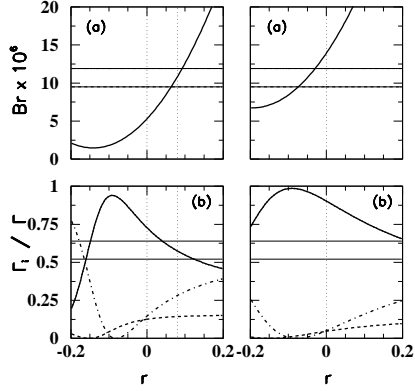}
\caption{Dependence of  branching ratio and polarization fractions
of $B^0 \to \phi K^{*0}$ on the long distance term. $B\to K^*$
form factors in \cite{Colangelo:1995jv} (left) and
\cite{Ball:2003rd} (right) are used in the factorized amplitude.
$r=0$ corresponds to  absence of rescattering.
 The three curves in (b)
refer to $f_L$ (continuous), $f_\perp$ (dashed) and $f_\parallel$
(dot-dashed). The horizontal lines represent the data for the
branching ratio (a) and for $f_L$ (b).}
 \label{fig:rate}
\end{figure}
 In ${\cal A}_{fact}$ we use the $B \to K^*$ form
factors
 in \cite{Colangelo:1995jv,Ball:2003rd},
with
 Wilson coefficients
 from \cite{Ali:1998eb}.
The result is shown in fig.\ref{fig:rate} \cite{Colangelo:2004rd}.
For the model \cite{Colangelo:1995jv},  $r\simeq 0.08$ gives the
experimental branching ratio and $f_L \simeq 0.55$. Using the form
factors in \cite{Ball:2003rd},  we reproduce the branching ratio
for $r\simeq-0.05$, though increasing $f_L$. However, this
conclusion
  depends on the value of the Wilson coefficients.  For
smaller
 $a_W$, in both cases a similar
long-distance contribution is required, reducing $f_L$.

Our conclusion  is  that rescattering can modify the helicity
amplitudes in penguin dominated  modes. On the other hand, such
effects are too small to affect  $B \to \rho \rho$ decays.
Actually, while  in the tree diagram in $B^0 \to \rho^+ \rho^-$
the CKM factor ($V^*_{ub}V_{ud}$) has similar size to that in the
rescattering diagrams ($V^*_{cb}V_{cd}$), the Wilson coefficient
in current-current transition is ${\cal O}(1)$. We expect to
observe FSI effects in colour-suppressed and other penguin induced
$B\to VV$ decays. For $B^+ \to \rho^0 K^{*+}$, including the
rescattering term, we get
 $f_L \simeq 0.7$, i.e. smaller (though
compatible within 2-$\sigma$) than the datum in Table
\ref{tab:tab2}. Hence,  our approach   can give a small $f_L(B \to
\phi K^*)$  at the price of  a smaller $f_L(B \to \rho K^*)$. For
$B^+ \to K^{*0} \rho^+$ very recent measurements  reported:
$f_L=0.79 \pm 0.08 \pm 0.04 \pm 0.02$ \cite{new1},
$f_L=0.50\pm0.19 \pm^{0.05}_{0.07}$ \cite{new2}.

Other analyses of non factorizable effects have been proposed
\cite{cheng,Kagan:2004uw}; they will not be discussed here. It is
only worth mentioning the common conclusion that there seems to be
no need of non standard mechanisms to understand the polarization
fractions in $B \to VV$ transitions, even though more refined
studies are required.

\vspace{0.4cm}
  I thank P. Colangelo and T.N. Pham for collaboration.
  Partial support from the EC Contract No.
HPRN-CT-2002-00311 (EURIDICE) is acknowledged.


\bibliographystyle{aipproc}   
\bibliographystyle{aipprocl} 



\end{document}